\documentclass[11pt,a4paper,hyper]{JHEP3}
\usepackage{feynarts,epsfig}

\newlength{\largfig}
\def\beq{\begin{equation}} 
\def\eeq{\end{equation}} 
 
\def\beqn{\begin{eqnarray}} 
\def\eeqn{\end{eqnarray}}

\def\Re{{\rm Re}}  
\def\Im{{\rm Im}}

\newcount\minutes 
\newcount\scratch 
 
\def\timestamp{%
\scratch=\time 
\divide\scratch by 60 
\edef\hours{\the\scratch} 
\multiply\scratch by 60 
\minutes=\time 
\advance\minutes by -\scratch 
---$\,$\hours:\null 
\ifnum\minutes< 10 0\fi 
\the\minutes} 

\title{Next-to-leading order QCD corrections to light Higgs Pair production via 
vector boson fusion}

\author{Terrance Figy \\
Institute of Particle Physics Phenomenology,
Durham University, Durham, DH1 3LE, United Kingdom \\
E-mail: \email{terrance.figy@durham.ac.uk}}
\abstract{
We present the NLO QCD corrections for light Higgs pair production via vector boson fusion 
at the LHC within the CP conserving type II two higgs doublet model in the form of 
a fully flexible parton--level Monte Carlo program. Scale dependences on integrated cross 
sections and distributions are reduced with QCD $K$--factors of order unity. 
}

\preprint{\arXivid{0806.2200}\\
IPPP/08/44 \\ 
DCPT/08/88 }
\begin{document}


\section{Introduction}
\label{sec:intro}
One of the primary goals of the CERN Large Hadron Collider (LHC) is a thorough investigation of
the mechanism of electroweak (EW) symmetry breaking and, more specifically, the 
discovery of one or more Higgs bosons and the determination of their properties~\cite{ATLAS,CMS}. 
In this context, vector-boson fusion (VBF) has emerged as 
particularly interesting  class of processes. Higgs boson production in VBF, i.e.\ the EW
reaction $qq\,\to\, qqh^{0}$, where the Higgs decay products are detected in
association with two tagging jets, offers a promising
discovery channel~\cite{Rainwater:1999gg} and, once its existence  
has been verified, will help to constrain the couplings of the Higgs
bosons to gauge bosons and fermions~\cite{Zeppenfeld:2000td}. 
QCD corrections to the total cross section for single Higgs boson production via VBF have been computed using the structure function approach \cite{Han:1992hr}. Distributions at NLO accuracy have recently become available through fully flexible parton--level Monte Carlo programs such as MCFM and VBFNLO~\cite{Figy:2003nv, Berger:2004pca,Ciccolini:2007jr}.

The two--Higgs doublet model (THDM) predicts the existence of two neutral CP--even Higgs bosons, 
one neutral CP--odd Higgs boson, and two charged Higgs bosons which have selfcouplings as well as couplings to gauge bosons and fermions~\cite{THDM,Gunion:2002zf}. Studies have shown that Higgs pair production at the LHC can serve 
as a probe of the Higgs potential~\cite{Moretti:2004dg,Moretti:2004wa,Boudjema:2001ii}. Of interest recently has been the process $pp \to h^{0}h^{0}jj \to b \bar{b} b \bar{b}jj$ via VBF in the context of the 
two higgs doublet model~\cite{Moretti:2007ca,Eboli:1987dy,Dobrovolskaya:1990kx,Abbasabadi:1988bk,Djouadi:1999rca}.  It was shown in Ref.~\cite{Moretti:2007ca} that in favorable THDM scenarios that it may be possible to extract the $H^{0} \to h^{0}h^{0} \to 4b$ resonance, thereby making the measurement of the trilinear $h^{0}h^{0}H^{0}$ coupling possible at the LHC. Assuming such a favorable scenario, the knowledge of QCD radiative corrections for this process will be needed in order to reduce the theoretical uncertainty on the total cross section and distributions. It is the 
aim of this paper to present the NLO QCD corrections for the process $pp \to h^{0}h^{0}jj \to 
b \bar{b} b \bar{b}jj$ via VBF in the form of a fully flexible partonic 
Monte Carlo program within the type II--CP conserving two higgs doublet model.

In Section~\ref{sec:thdm} we give an overview of the THDM and establish benchmark points used in our 
simulations. Section~\ref{sec:calc} lays out the 
details of the NLO calculation. Cross sections and distributions for the LHC are given in Section~\ref{sec:res}. Conclusions are given in Section~\ref{sec:concl}.

\section{Two Higgs Doublet Model Parameters}
\label{sec:thdm}
The THDM contains two $SU(2)$ doublets, $\Phi_{1}$ and $\Phi_{2}$, of weak hypercharge
$Y=1$. In the CP conserving THDM there is freedom in the choice of the Higgs boson--fermion
couplings \cite{THDM}: type I, in which only one Higgs doublet couples to fermions; and type II,
in which the neutral member of one Higgs doublet couples to up-type quarks and the neutral member 
of the other Higgs doublet couples to down-type quarks and leptons. Flavor changing neutral currents
(FCNC) mediated by the Higgs bosons are automatically absent in both type I and type II THDMs \cite{Glashow:1976nt}. In this work, we only consider type II Higgs boson--fermion couplings.
The most general THDM scalar potential which is 
invariant under $SU(2)_{L} \otimes U(1)_{Y}$ and conserves CP is given by \cite{THDM},
\beqn
\label{eq:pot}
V(\Phi_{1},\Phi_{2})& =& \lambda_{1} \left( \Phi_{1}^{\dagger} \Phi_{1} -v_{1}^{2} \right )^{2}
+ \lambda_{2} \left ( \Phi_{2}^{\dagger} \Phi_{2} -v_{2}^{2} \right )^{2}  \nonumber \\ 
&+& \lambda_{3} \left [(\Phi_{1}^{\dagger} \Phi_{1} -v_{1}^{2}) + 
(\Phi_{2}^{\dagger} \Phi_{2} -v_{2}^{2}) \right ]^{2}  \nonumber \\
&+&\lambda_{4} \left [\left ( \Phi_{1}^{\dagger} \Phi_{1} \right )\left ( \Phi_{2}^{\dagger} \Phi_{2} \right ) 
- \left ( \Phi_{1}^{\dagger} \Phi_{2} \right )\left ( \Phi_{1}^{\dagger} \Phi_{2} \right ) \right ]  \\ \nonumber 
&+& \lambda_{5} \left  [  \Re \left ( \Phi_{1}^{\dagger} \Phi_{2} \right )  - v_{1} v_{2}  \right ]^{2} 
+ \lambda_{6} \left [  \Im \left ( \Phi_{1}^{\dagger} \Phi_{2} \right ) \right ]^2 \, ,
\eeqn 
with two real parameters, $v_{1},v_{2}$ of mass dimension one and $6$ real dimensionless parameters, 
$\lambda_{1} ,\ldots, \lambda_{6}$. The minimum of the potential given by Eq.~(\ref{eq:pot}) occurs at 
$\Phi_{i}=(0, \frac{v_{i}}{\sqrt{2}})^{T}$ 
for $(i=1,2)$. The physical spectrum of the Higgs sector of the CP-conserving THDM consists of two neutral CP even 
Higgs bosons, ($h^{0},H^{0}$) , one neutral CP-odd Higgs boson, $A^{0}$, 
and two charged Higgs bosons, $(H^{+}, H^{-})$. The $8$ parameters of the Higgs sector can also be taken as the vacuum expectation value, $v=\sqrt{v_{1}^{2} + v_{2}^{2}}=\sqrt{(\sqrt{2} G_{F})^{-1}}$, 
the masses of the Higgs bosons, $m_{h^{0}}, m_{H^{0}}, m_{A^{0}}$ and $m_{H^{\pm}}$, the mixing angles $\alpha$ and $\beta$, and $\lambda_{5}$. 

The parameters of the Higgs potential, Eq.~(\ref{eq:pot}), can be restricted by imposing theoretical
requirements for the consistency of the model.  We use, both, the requirement of vacuum stability \cite{VAC}
and perturbative unitarity \cite{PERUNTY} for the tree--level coupling constants. The condition of 
vacuum stability is given by Eq.~(2) of Ref.~\cite{Asakawa:2005nx}.  Perturbative unitarity requires that the 
magnitudes of all tree--level S--wave amplitudes for elastic scattering of longitudinally polarized gauge 
and Higgs bosons stay in the limit set by unitarity. Here, we consider the $14$ neutral channels
of Ref.~\cite{Kanemura:1993hm}. $\rho$ parameter constraints from electroweak precision data have also been considered
\cite{RHO,Yao:2006px}.

In this paper, we will consider two benchmark points tabulated in Table~\ref{tbl:thdm} which in fact
satisfy the above mentioned requirements of vacuum stability and perturbative unitarity. 
Benchmark point $B1$ corresponds to a scenario in which the light Higgs boson does not couple to gauge bosons, i.e, $\cos(\alpha-\beta)=1$. In this scenario it will not be possible to produce a single light Higgs via VBF since the $h^{0} VV$ coupling is zero. Benchmark point $B2$ is scenario in which the heavy Higgs $H^{0}$ decouples from the gauge bosons, i.e., $\sin(\alpha-\beta)=1$. 

\TABLE{
\label{tbl:thdm}
\caption{THDM benchmark points.}
\begin{tabular}{|c|c|c|c|c|c|c|c|}
 \hline
           & $ \sin \alpha$ & $\tan \beta$ & $\lambda_{5}$ & $m_{A^{0}}$ & $m_{h^{0}}$ & $m_{H^{0}}$ & $m_{H^{\pm}}$  \\
 $B1$      & $0.832$        & $1.50$       & $-3.50$       & $295$ GeV   & $120$ GeV   & $300$ GeV   & $385$ GeV \\
 $B2$      & $-0.554$        & $1.50$       & $0$       & $295$ GeV   & $120$ GeV   & $300$ GeV   & $385$ GeV \\
\hline
\end{tabular}
}

\section{The NLO Calculation}\label{sec:calc}
At leading order (LO), light Higgs pair production via VBF can, effectively, 
be viewed (see Fig.~\ref{fig:brn}$a$) as the elastic scattering of two (anti)quarks, mediated by $t$--channel $W$ or $Z$ exchange, with two light Higgs boson radiated off the weak-boson propagator.
The ``blobs" in Fig.~\ref{fig:brn} for the process $\bar{q}Q \to \bar{q}Q h^{0}h^{0}$ represent the
vector boson scattering processes $W^{+}W^{-} \to h^{0}h^{0}$ (see Fig.~\ref{fig:wwhh}) and 
$ZZ \to h^{0}h^{0}$ (see Fig.~\ref{fig:zzhh}), for charged current (CC) and neutral current (NC)
processes, respectively.  The generalization to crossed processes ($\bar{q} \to q$ 
and/or $Q \to \bar{Q}$) is straightforward.  In principal we should consider the double higgstrahlung
process $Vh^{0}h^{0}$ with $V \to q \bar{q}$ and the exchange of identical fermions in the initial or final state.  
However, in phase space regions with widely separated quarks jets of high invariant mass, 
the interference of these additional graphs is strongly suppressed by the large momentum transfer
in the weak-boson propagators.  Color suppression further makes these effects negligible. We, therefore,
treat double higgstrahlung as a separate process and systematically neglect any identical particle effects
as in the case of single Higgs boson production via VBF \cite{cogeorg,Ciccolini:2007jr}. Further, gluon
fusion $h^{0} h^{0}jj$ production is treated as a separate process, since, the $\mathcal{O}(\alpha^2 \alpha_{s}^{2})$ corrections are on the order of atobarns~\cite{Andersen:2007mp,Bredenstein:2008tm}.

The NLO calculation is performed in complete analogy to Ref.~\cite{Figy:2003nv}.  The real emission 
graphs can be obtained by attaching the gluon to the quark lines of Fig.~\ref{fig:brn}$a$ in all 
possible ways. Two distinct non-interfering color structures result: Feynman graphs with a single gluon
attached to the upper quark line and Feynman graphs with single gluon attached to the lower quark line. 
Gluon initiated processes are obtained by crossing the final state gluon with the initial state (anti)quark.
The result are graphs with $t$--channel and $s$--channel weak boson exchange. 
For consistency of the calculation we neglect the $s$--channel process $gq \to Vh^{0}h^{0}q$, since, we have neglected double higgstrahlung at LO.

All amplitudes are calculated numerically, using the helicity-amplitude formalism of Ref.~\cite{HZ}.
Matrix elements for VBF proceses take the general form,
\beqn
\mathcal{M}^{\bar{q}Q} \sim T_{VV}^{\mu \nu} J_{\mu}^{\bar{q}} J_{\nu}^{Q},
\eeqn
with $J^{\bar{q}}_{\mu}$ and $J^{Q}_{\mu}$ being the two quark currents shown in Fig.~\ref{fig:brn}. 
We have used the THDM implementation of FeynArts and FormCalc \cite{FORMCALC,FEYNARTS} to 
generate model predictions for the tensors $T_{WW}^{\mu \nu}$  and $T_{ZZ}^{\mu \nu}$ for CC and NC vector boson
fusion processes, respectively, for the case of off-shell vector bosons. 
We have introduced finite widths for $h^{0}$ and $H^{0}$ in the $s$--channel Higgs propagators.
The subsequent decay of the light Higgs boson $h^{0}$ to $b \bar{b}$ is performed within the 
narrow--width approximation.

Divergences arising from the real corrections are regulated in $d=4-2 \epsilon$ spacetime dimensions by
using the \emph{Catani-Seymour} dipole subtraction method \cite{CS} in the dimensional reduction
scheme \cite{Siegel:1979wq}. Formulas for the subtraction terms and finite collinear pieces are identical
to the ones for single Higgs production via VBF and are given in Ref.~\cite{Figy:2003nv}.

\FIGURE{
\epsfig{file=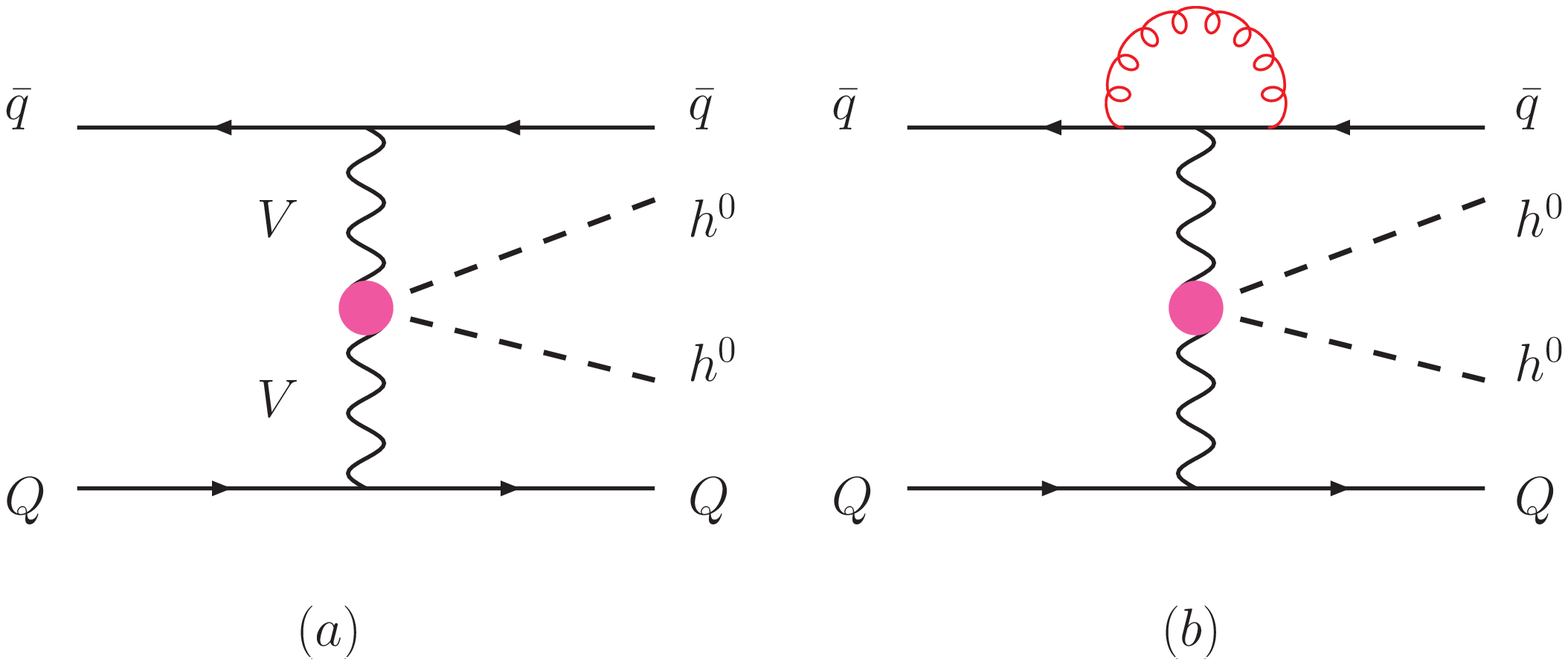,scale=0.5} 
\caption{Feynman graphs contributing to $\bar{q}Q \to \bar{q}Qh^{0}h^{0}$ 
at ($a$) tree-level and ($b$) including virtual corrections to the upper line.  
The ``blobs" correspond to $VV  h^{0}h^{0}$  effective vertices which are 
represented by the tensors $T_{VV}^{\mu \nu}$ where $V=Z,W^{\pm}$.}\label{fig:brn}
}

\FIGURE{
\unitlength=0.8bp%

\begin{feynartspicture}(432,504)(3,3.3)

\FADiagram{($a$)}
\FAProp(0.,15.)(10.,10.)(0.,){/Sine}{1}
\FALabel(4.78682,11.5936)[tr]{$W^{+}$}
\FAProp(0.,5.)(10.,10.)(0.,){/Sine}{-1}
\FALabel(5.21318,6.59364)[tl]{$W^{+}$}
\FAProp(20.,15.)(10.,10.)(0.,){/ScalarDash}{0}
\FALabel(14.8986,13.1828)[br]{$h^0$}
\FAProp(20.,5.)(10.,10.)(0.,){/ScalarDash}{0}
\FALabel(15.1014,8.18276)[bl]{$h^0$}
\FAVert(10.,10.){0}

\FADiagram{($b$)}
\FAProp(0.,15.)(6.,10.)(0.,){/Sine}{1}
\FALabel(2.48771,11.7893)[tr]{$W^{+}$}
\FAProp(0.,5.)(6.,10.)(0.,){/Sine}{-1}
\FALabel(3.51229,6.78926)[tl]{$W^{+}$}
\FAProp(20.,15.)(14.,10.)(0.,){/ScalarDash}{0}
\FALabel(16.6478,13.0187)[br]{$h^0$}
\FAProp(20.,5.)(14.,10.)(0.,){/ScalarDash}{0}
\FALabel(17.3522,8.01869)[bl]{$h^0$}
\FAProp(6.,10.)(14.,10.)(0.,){/ScalarDash}{0}
\FALabel(10.,9.18)[t]{$h^0$}
\FAVert(6.,10.){0}
\FAVert(14.,10.){0}

\FADiagram{($c$)}
\FAProp(0.,15.)(6.,10.)(0.,){/Sine}{1}
\FALabel(2.48771,11.7893)[tr]{$W^{+}$}
\FAProp(0.,5.)(6.,10.)(0.,){/Sine}{-1}
\FALabel(3.51229,6.78926)[tl]{$W^{+}$}
\FAProp(20.,15.)(14.,10.)(0.,){/ScalarDash}{0}
\FALabel(16.6478,13.0187)[br]{$h^0$}
\FAProp(20.,5.)(14.,10.)(0.,){/ScalarDash}{0}
\FALabel(17.3522,8.01869)[bl]{$h^0$}
\FAProp(6.,10.)(14.,10.)(0.,){/ScalarDash}{0}
\FALabel(10.,9.18)[t]{$H^0$}
\FAVert(6.,10.){0}
\FAVert(14.,10.){0}

\FADiagram{($d$)}
\FAProp(0.,15.)(10.,14.)(0.,){/Sine}{1}
\FALabel(4.84577,13.4377)[t]{$W^{+}$}
\FAProp(0.,5.)(10.,6.)(0.,){/Sine}{-1}
\FALabel(5.15423,4.43769)[t]{$W^{+}$}
\FAProp(20.,15.)(10.,14.)(0.,){/ScalarDash}{0}
\FALabel(14.8706,15.3135)[b]{$h^0$}
\FAProp(20.,5.)(10.,6.)(0.,){/ScalarDash}{0}
\FALabel(15.1294,6.31355)[b]{$h^0$}
\FAProp(10.,14.)(10.,6.)(0.,){/ScalarDash}{1}
\FALabel(8.93,10.)[r]{$H^{+}$}
\FAVert(10.,14.){0}
\FAVert(10.,6.){0}

\FADiagram{($e$)}
\FAProp(0.,15.)(10.,14.)(0.,){/Sine}{1}
\FALabel(4.84577,13.4377)[t]{$W^{+}$}
\FAProp(0.,5.)(10.,6.)(0.,){/Sine}{-1}
\FALabel(5.15423,4.43769)[t]{$W^{+}$}
\FAProp(20.,15.)(10.,14.)(0.,){/ScalarDash}{0}
\FALabel(14.8706,15.3135)[b]{$h^0$}
\FAProp(20.,5.)(10.,6.)(0.,){/ScalarDash}{0}
\FALabel(15.1294,6.31355)[b]{$h^0$}
\FAProp(10.,14.)(10.,6.)(0.,){/ScalarDash}{1}
\FALabel(8.93,10.)[r]{$G^{+}$}
\FAVert(10.,14.){0}
\FAVert(10.,6.){0}

\FADiagram{($f$)}
\FAProp(0.,15.)(10.,14.)(0.,){/Sine}{1}
\FALabel(4.84577,13.4377)[t]{$W^{+}$}
\FAProp(0.,5.)(10.,6.)(0.,){/Sine}{-1}
\FALabel(5.15423,4.43769)[t]{$W^{+}$}
\FAProp(20.,15.)(10.,14.)(0.,){/ScalarDash}{0}
\FALabel(14.8706,15.3135)[b]{$h^0$}
\FAProp(20.,5.)(10.,6.)(0.,){/ScalarDash}{0}
\FALabel(15.1294,6.31355)[b]{$h^0$}
\FAProp(10.,14.)(10.,6.)(0.,){/Sine}{1}
\FALabel(8.93,10.)[r]{$W^{+}$}
\FAVert(10.,14.){0}
\FAVert(10.,6.){0}

\FADiagram{($g$)}
\FAProp(0.,15.)(10.,14.)(0.,){/Sine}{1}
\FALabel(4.84577,13.4377)[t]{$W^{+}$}
\FAProp(0.,5.)(10.,6.)(0.,){/Sine}{-1}
\FALabel(5.15423,4.43769)[t]{$W^{+}$}
\FAProp(20.,15.)(10.,6.)(0.,){/ScalarDash}{0}
\FALabel(16.98,13.02)[br]{$h^0$}
\FAProp(20.,5.)(10.,14.)(0.,){/ScalarDash}{0}
\FALabel(17.52,8.02)[bl]{$h^0$}
\FAProp(10.,14.)(10.,6.)(0.,){/ScalarDash}{1}
\FALabel(9.03,10.)[r]{$H^{+}$}
\FAVert(10.,14.){0}
\FAVert(10.,6.){0}

\FADiagram{($h$)}
\FAProp(0.,15.)(10.,14.)(0.,){/Sine}{1}
\FALabel(4.84577,13.4377)[t]{$W^{+}$}
\FAProp(0.,5.)(10.,6.)(0.,){/Sine}{-1}
\FALabel(5.15423,4.43769)[t]{$W^{+}$}
\FAProp(20.,15.)(10.,6.)(0.,){/ScalarDash}{0}
\FALabel(16.98,13.02)[br]{$h^0$}
\FAProp(20.,5.)(10.,14.)(0.,){/ScalarDash}{0}
\FALabel(17.52,8.02)[bl]{$h^0$}
\FAProp(10.,14.)(10.,6.)(0.,){/ScalarDash}{1}
\FALabel(9.03,10.)[r]{$G^{+}$}
\FAVert(10.,14.){0}
\FAVert(10.,6.){0}

\FADiagram{($i$)}
\FAProp(0.,15.)(10.,14.)(0.,){/Sine}{1}
\FALabel(4.84577,13.4377)[t]{$W^{+}$}
\FAProp(0.,5.)(10.,6.)(0.,){/Sine}{-1}
\FALabel(5.15423,4.43769)[t]{$W^{+}$}
\FAProp(20.,15.)(10.,6.)(0.,){/ScalarDash}{0}
\FALabel(16.98,13.02)[br]{$h^0$}
\FAProp(20.,5.)(10.,14.)(0.,){/ScalarDash}{0}
\FALabel(17.52,8.02)[bl]{$h^0$}
\FAProp(10.,14.)(10.,6.)(0.,){/Sine}{1}
\FALabel(9.03,10.)[r]{$W^{+}$}
\FAVert(10.,14.){0}
\FAVert(10.,6.){0}
\end{feynartspicture}

\caption{Feynman graphs for the process $W^{+}W^{-} \to h^{0}h^{0}$.}
\label{fig:wwhh}
}

\FIGURE{
\unitlength=0.8bp%

\begin{feynartspicture}(432,504)(3,3.3)

\FADiagram{($a$)}
\FAProp(0.,15.)(10.,10.)(0.,){/Sine}{0}
\FALabel(4.78682,11.5936)[tr]{$Z$}
\FAProp(0.,5.)(10.,10.)(0.,){/Sine}{0}
\FALabel(5.21318,6.59364)[tl]{$Z$}
\FAProp(20.,15.)(10.,10.)(0.,){/ScalarDash}{0}
\FALabel(14.8986,13.1828)[br]{$h^0$}
\FAProp(20.,5.)(10.,10.)(0.,){/ScalarDash}{0}
\FALabel(15.1014,8.18276)[bl]{$h^0$}
\FAVert(10.,10.){0}

\FADiagram{($b$)}
\FAProp(0.,15.)(6.,10.)(0.,){/Sine}{0}
\FALabel(2.48771,11.7893)[tr]{$Z$}
\FAProp(0.,5.)(6.,10.)(0.,){/Sine}{0}
\FALabel(3.51229,6.78926)[tl]{$Z$}
\FAProp(20.,15.)(14.,10.)(0.,){/ScalarDash}{0}
\FALabel(16.6478,13.0187)[br]{$h^0$}
\FAProp(20.,5.)(14.,10.)(0.,){/ScalarDash}{0}
\FALabel(17.3522,8.01869)[bl]{$h^0$}
\FAProp(6.,10.)(14.,10.)(0.,){/ScalarDash}{0}
\FALabel(10.,9.18)[t]{$h^0$}
\FAVert(6.,10.){0}
\FAVert(14.,10.){0}

\FADiagram{($c$)}
\FAProp(0.,15.)(6.,10.)(0.,){/Sine}{0}
\FALabel(2.48771,11.7893)[tr]{$Z$}
\FAProp(0.,5.)(6.,10.)(0.,){/Sine}{0}
\FALabel(3.51229,6.78926)[tl]{$Z$}
\FAProp(20.,15.)(14.,10.)(0.,){/ScalarDash}{0}
\FALabel(16.6478,13.0187)[br]{$h^0$}
\FAProp(20.,5.)(14.,10.)(0.,){/ScalarDash}{0}
\FALabel(17.3522,8.01869)[bl]{$h^0$}
\FAProp(6.,10.)(14.,10.)(0.,){/ScalarDash}{0}
\FALabel(10.,9.18)[t]{$H^0$}
\FAVert(6.,10.){0}
\FAVert(14.,10.){0}

\FADiagram{($d$)}
\FAProp(0.,15.)(10.,14.)(0.,){/Sine}{0}
\FALabel(4.84577,13.4377)[t]{$Z$}
\FAProp(0.,5.)(10.,6.)(0.,){/Sine}{0}
\FALabel(5.15423,4.43769)[t]{$Z$}
\FAProp(20.,15.)(10.,14.)(0.,){/ScalarDash}{0}
\FALabel(14.8706,15.3135)[b]{$h^0$}
\FAProp(20.,5.)(10.,6.)(0.,){/ScalarDash}{0}
\FALabel(15.1294,6.31355)[b]{$h^0$}
\FAProp(10.,14.)(10.,6.)(0.,){/ScalarDash}{0}
\FALabel(9.18,10.)[r]{$A^0$}
\FAVert(10.,14.){0}
\FAVert(10.,6.){0}

\FADiagram{($e$)}
\FAProp(0.,15.)(10.,14.)(0.,){/Sine}{0}
\FALabel(4.84577,13.4377)[t]{$Z$}
\FAProp(0.,5.)(10.,6.)(0.,){/Sine}{0}
\FALabel(5.15423,4.43769)[t]{$Z$}
\FAProp(20.,15.)(10.,14.)(0.,){/ScalarDash}{0}
\FALabel(14.8706,15.3135)[b]{$h^0$}
\FAProp(20.,5.)(10.,6.)(0.,){/ScalarDash}{0}
\FALabel(15.1294,6.31355)[b]{$h^0$}
\FAProp(10.,14.)(10.,6.)(0.,){/ScalarDash}{0}
\FALabel(9.18,10.)[r]{$G^0$}
\FAVert(10.,14.){0}
\FAVert(10.,6.){0}

\FADiagram{($f$)}
\FAProp(0.,15.)(10.,14.)(0.,){/Sine}{0}
\FALabel(4.84577,13.4377)[t]{$Z$}
\FAProp(0.,5.)(10.,6.)(0.,){/Sine}{0}
\FALabel(5.15423,4.43769)[t]{$Z$}
\FAProp(20.,15.)(10.,14.)(0.,){/ScalarDash}{0}
\FALabel(14.8706,15.3135)[b]{$h^0$}
\FAProp(20.,5.)(10.,6.)(0.,){/ScalarDash}{0}
\FALabel(15.1294,6.31355)[b]{$h^0$}
\FAProp(10.,14.)(10.,6.)(0.,){/Sine}{0}
\FALabel(8.93,10.)[r]{$Z$}
\FAVert(10.,14.){0}
\FAVert(10.,6.){0}

\FADiagram{($g$)}
\FAProp(0.,15.)(10.,14.)(0.,){/Sine}{0}
\FALabel(4.84577,13.4377)[t]{$Z$}
\FAProp(0.,5.)(10.,6.)(0.,){/Sine}{0}
\FALabel(5.15423,4.43769)[t]{$Z$}
\FAProp(20.,15.)(10.,6.)(0.,){/ScalarDash}{0}
\FALabel(16.98,13.02)[br]{$h^0$}
\FAProp(20.,5.)(10.,14.)(0.,){/ScalarDash}{0}
\FALabel(17.52,8.02)[bl]{$h^0$}
\FAProp(10.,14.)(10.,6.)(0.,){/ScalarDash}{0}
\FALabel(9.28,10.)[r]{$A^0$}
\FAVert(10.,14.){0}
\FAVert(10.,6.){0}

\FADiagram{($h$)}
\FAProp(0.,15.)(10.,14.)(0.,){/Sine}{0}
\FALabel(4.84577,13.4377)[t]{$Z$}
\FAProp(0.,5.)(10.,6.)(0.,){/Sine}{0}
\FALabel(5.15423,4.43769)[t]{$Z$}
\FAProp(20.,15.)(10.,6.)(0.,){/ScalarDash}{0}
\FALabel(16.98,13.02)[br]{$h^0$}
\FAProp(20.,5.)(10.,14.)(0.,){/ScalarDash}{0}
\FALabel(17.52,8.02)[bl]{$h^0$}
\FAProp(10.,14.)(10.,6.)(0.,){/ScalarDash}{0}
\FALabel(9.28,10.)[r]{$G^0$}
\FAVert(10.,14.){0}
\FAVert(10.,6.){0}

\FADiagram{($i$)}
\FAProp(0.,15.)(10.,14.)(0.,){/Sine}{0}
\FALabel(4.84577,13.4377)[t]{$Z$}
\FAProp(0.,5.)(10.,6.)(0.,){/Sine}{0}
\FALabel(5.15423,4.43769)[t]{$Z$}
\FAProp(20.,15.)(10.,6.)(0.,){/ScalarDash}{0}
\FALabel(16.98,13.02)[br]{$h^0$}
\FAProp(20.,5.)(10.,14.)(0.,){/ScalarDash}{0}
\FALabel(17.52,8.02)[bl]{$h^0$}
\FAProp(10.,14.)(10.,6.)(0.,){/Sine}{0}
\FALabel(9.03,10.)[r]{$Z$}
\FAVert(10.,14.){0}
\FAVert(10.,6.){0}
\end{feynartspicture}

\caption{Feynman graphs for the process $Z^{0}Z^{0} \to h^{0}h^{0}$.}
\label{fig:zzhh}}

\section{Results for the LHC}
\label{sec:res}
The goal of our calculation is a precise prediction of the LHC cross
section for light Higgs boson pair production in VBF with two or more jets in 
the context of the THDM. In order to reconstruct jets from the final-state partons,
the $k_T$ algorithm~\cite{kToriginal} as described in Ref.~\cite{kTrunII} is used, with resolution parameter $D=0.8$. 
These jets are required to have 
\begin{eqnarray}
\label{cuts1}
p_{Tj} \geq 20~{\rm GeV} \, , \qquad\qquad |y_j| \leq 4.5 \, .
\end{eqnarray}
Here $y_j$ denotes the rapidity of the (massive) jet momentum which is 
reconstructed as the four-vector sum of massless partons of 
pseudorapidity $|\eta|<5$. 

At LO, there are exactly two massless final state partons.  The two
hardest jets are identified as tagging jets, provided they pass the $k_T$
algorithm and the cuts described above. At NLO these jets may be
composed of two partons (recombination effect) or three well-separated
partons may be encountered, of which at least two 
satisfy the cuts of Eq.~(\ref{cuts1}) and would give rise to either  
two or three-jet events. As with LHC data, a choice needs
to be made for selecting the tagging jets in such a multijet situation.
Here the ``$p_{T}$-method" is chosen.  For a given event, the tagging
jets are defined as the two jets with the highest transverse momentum
with 
\beq\label{eq:tagcuts}
p_{Tj}^{\rm tag} \ge 30 ~{\rm GeV}, \quad \quad |y_{j}^{\rm tag}| \le 4.5.
\eeq
$b$ --jets arising from decays of the two light Higgs bosons ($h^{0} \to b \bar{b}$), are restricted by the 
following cuts: 
\begin{eqnarray}
p_{T b} \geq 30~{\rm GeV} \,,\qquad |\eta_{b}| \leq 2.5  \,,\qquad 
\triangle R_{jb} \geq 0.6 \, ,\qquad  \triangle R_{bb} \geq 0.7 
\end{eqnarray} 
where $\triangle R_{jb}$ denotes the jet-$b$ separation in the
rapidity-azimuthal angle plane. In addition, the $b$--jets are
required to fall between the two tagging jets in rapidity: 
\begin{eqnarray}
y_{j,{\rm min}}^{\rm tag}<\eta_{b}<y_{j,{\rm max}}^{\rm tag} \, .
\end{eqnarray} 

Backgrounds to VBF are significantly suppressed by
requiring a large rapidity separation for the two tagging jets. 
Tagging jets are required to reside in opposite detector hemispheres with
\beq \label{eq:ophem}
y_{j}^{\rm tag ~1} \cdot y_{j}^{\rm tag ~2} < 0
\eeq
and to have a large rapidity separation of 
\begin{eqnarray}\label{eq:yjjcut}
\Delta y_{jj} = | y_{j}^{\rm tag ~1}-y_{j}^{\rm tag ~2}|>4\;,
\end{eqnarray}
sometimes called ``rapidity gap cut''. 
QCD backgrounds can be reduced by  imposing a
lower bound on the invariant mass of the tagging jets of 
\begin{eqnarray} \label{eq:dijet}
 m_{jj} =\sqrt{ (p_{j}^{\rm tag~1}+p_{j}^{\rm tag~2})^2} > 600 ~{\rm GeV}.
\end{eqnarray}

In all subsequent calculations we use the input parameters for defining 
Standard Model (SM) couplings as listed in Table \ref{tbl:smparm}.  
Other SM couplings are computed using LO
electroweak relations. Cross sections are computed using CTEQ6M parton
distributions \cite{cteq6} with $\alpha_s(M_Z)=0.118$ for all NLO results and CTEQ6L1 parton
distributions with $\alpha_s(M_Z)=0.130$ for all leading order cross sections.  The running of the
strong coupling is evaluated at two-loop order for all NLO results. 
In the following we use benchmark point $B1$ for NLO and LO Monte Carlo simulations. 

\TABLE[tb]{
\caption{SM input parameters}
\begin{tabular}{ccc}
\hline
\hline
 $M_{Z}$  &  $M_{W}$ & $G_{F}$ \\
 $91.188~{\rm GeV}$ & $80.416~{\rm GeV}$ & $1.16639 \times 10^{-5}/{\rm GeV}^2$  \\ 
\hline
\hline
\end{tabular}%
\label{tbl:smparm}}

In Figures~\ref{fig:sigxi1} and \ref{fig:sigxi2}, we show the scale dependence of the total cross section within the cuts of Eqs. (\ref{cuts1})-(\ref{eq:dijet}) for the process $pp \to h^{0}h^{0} jj \to b \bar{b} b \bar{b} jj$ via VBF at the LHC. Scale variations are shown for (a) LO results with $\mu_{F}=\xi \mu_{0}$ (black dotted line), (b) NLO results with $\mu_{F} = \xi \mu_{0}$ and $\mu_{R} = \mu_{0}$ (blue dot--dashed line), 
(c) NLO results with $\mu_{R} = \xi \mu_{0}$ and $\mu_{F} = \mu_{0}$ (green dashed line), and (d)
LO results with $\mu_{F}=\xi \mu_{0}$ (red solid line). 
In Figure~\ref{fig:sigxi1}, we choose a fixed reference scale, 
$\mu_{0} = 2 m_{h}$.  In Figure~\ref{fig:sigxi2}, we choose the reference scale to be the virtuality of the 
exchanged $t$--channel vector boson, $q_{V}$, shown in Figure~\ref{fig:brn}. 
For $\mu_{F}=\xi \mu_{0}$ with $0.2<\xi<5$ the scale variation of the LO cross section is
$+19 \%$ to $-14 \%$ for both choices of reference scale $\mu_{0}$.  While at NLO, the 
scale variations are $-1.34 \% $ to $-2.64 \%$ for $\mu_{F}=\mu_{R} = 5^{\pm 1} (2 m_{h})$ and 
$-5 \%$ to $0.5 \%$ for $\mu_{F}=\mu_{R} = 5^{\pm 1} q_{V}$. Table~\ref{tbl:xsec} lists 
total cross sections at NLO for the benchmark scenarios of Table~\ref{tbl:thdm}. For benchmark $B2$ the 
cross section is below $1$ femtobarn due to the fact that the heavy Higgs boson does not couple to gauge bosons.  

Our Monte Carlo program allows the analysis of arbitrary infrared and collinear
safe distributions with NLO QCD accuracy. In order to assess the impact of 
the NLO corrections we compare LO and NLO results by plotting the dynamical 
$K$ factor
\beq
K(x) = \frac{d \sigma_{3}^{NLO}(\mu_{R} = \mu_{F}=\mu_{0})/dx}
{d \sigma_{3}^{LO}(\mu_R=\mu_F = \mu_{0})/dx}
\eeq
for our reference scale of $\mu_0$. The stability of the results 
is represented via the scale dependence, given by the ratio of cross sections,
\beq
R(x) = \frac{d \sigma_{3}(\mu_R=\mu_F=\xi \mu_{0})/dx}
{d \sigma_{3} (\mu_R=\mu_F=\mu_{0})/dx}\;.
\eeq 
We plot results for $\xi = 1/2$ and $2$ with $\mu_{0}=q_{V}$ for 
NLO and LO distributions.

\TABLE{
\label{tbl:xsec}
\begin{tabular}{|c|c|c|c|}
 \hline
Benchmark & $ \sigma_{\rm LO}({\rm fb})$ & $\sigma_{\rm NLO}({\rm fb})$ & $K$--factor  \\
\hline 
$B1$      & $  86.87$                    & $87.62$                      &$1.0086$ \\
$B2$      & $ 0.1864$                    & $0.1858$                     &$0.9969$\\
\hline
\end{tabular}
\caption{Integrated LO cross sections, $\sigma_{LO}$, NLO cross sections, $\sigma_{LO}$, in femtobarns (fb). The QCD  $K$--factors are defined as $K=\sigma_{NLO}/\sigma_{LO}$.  The renormalization and factorization scale has been set to $\mu_{R} = \mu_{F} = q_{V}$.}
}

Figure~\ref{fig:ptmax} shows the maximum transverse momentum of the two tagging jets defined as
$p_{T,tag}^{\rm max} = {\rm max}(p_{T}^{\rm tag~1},p_{T}^{\rm tag~2})$. The left panel shows the distribution at 
LO (dashed green histogram) and at NLO (solid red histogram). The right panel shows phase space dependent $K$ factor (solid green histogram). Scale variations $\mu_{R} = \mu_{F} = \xi \mu_{0}$ for $\xi = 0.5$ and $2$ are shown for LO (dotted histograms) and NLO (dashed histograms). 
While at LO the scale variations are $\pm 8 \% $, they are reduced to $\pm 1 \%$ at NLO.
Figure~\ref{fig:yjj} shows the rapidity separation of the two tagging jets. At NLO the tagging jet rapidity separation tends to increase which is typical behavior for VBF production processes \cite{Figy:2003nv}. 

Figure~\ref{fig:ptbmax} shows the distribution in the maximum transverse momentum of the four $b$--jets resulting from the decay of the light Higgs bosons defined as $p_{T,b}^{\rm max}={\rm max}(p_{T,b1},\ldots,p_{T,b4})$. The distribution is peaked in the vicinity of $100$ GeV because we have chosen the light Higgs boson mass to be $m_{h}=120$ GeV. A lower light Higgs mass would result in a softer $b$--jet. Figure~\ref{fig:m4b} shows the distribution in the four $b$--jet invariant mass, $m_{4b}=\sqrt{(p_{b1} + \cdots + p_{b4})^2}$. A peak occurs 
a around the mass of the heavy Higgs boson, $m_{H}=300$ GeV. In both distributions the scale dependence is reduced with relatively flat phase space dependent $K$ factors.

\FIGURE{
\epsfig{file=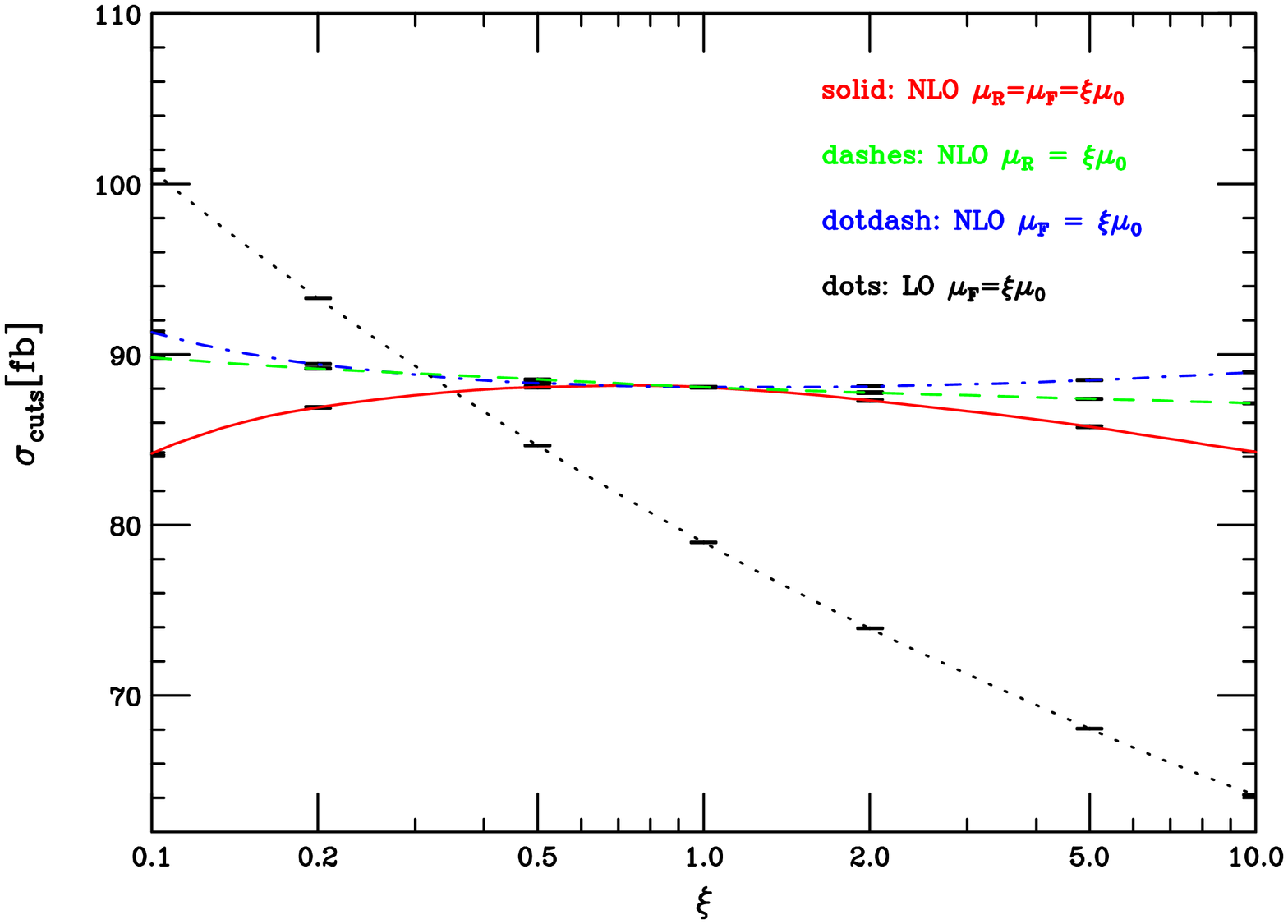,angle=90,width=4.5in}
\caption{Scale dependence of the total cross section at LO and NLO within 
the cuts of Eqs. (\ref{cuts1})-(\ref{eq:dijet}) for VBF $h^{0}h^{0}jj$ production
at the LHC. The factorization scale $\mu_F$ and the renormalization scale 
$\mu_R$ are taken as multiples, $\xi \mu_{0}$,  of the fixed reference 
scale $\mu_{0} = 2 m_{h}$. The NLO curves are  
for $\mu_R=\mu_F=\xi \mu_{0}$ (solid red line), 
$\mu_F = \mu_{0}$ and $\mu_R=\xi \mu_{0}$ (dashed green line),
and $\mu_F=\xi \mu_{0}$ and $\mu_R = \xi \mu_{0}$ (dot-dashed blue line ).  
The dotted black curve shows the scale dependence of the LO cross section 
for $\mu_F = \xi \mu_0$.}%
\label{fig:sigxi1}
} 

\FIGURE{
\epsfig{file=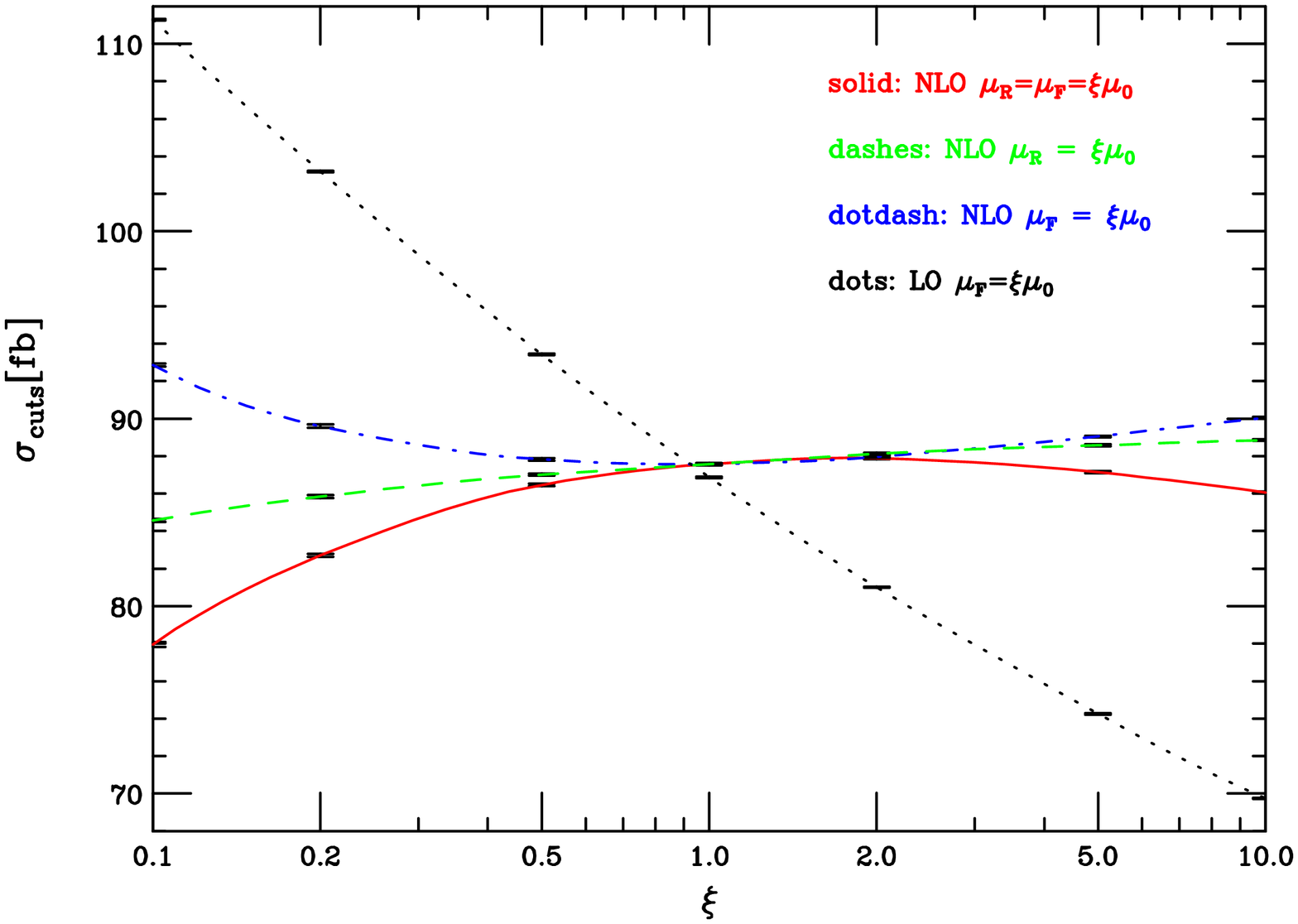,angle=90,width=4.5in}
\caption{The caption is the same as in Fig.~\ref{fig:sigxi1} except $\mu_{0}=q_{V}$.}
\label{fig:sigxi2}
} 

\FIGURE{
\epsfig{file=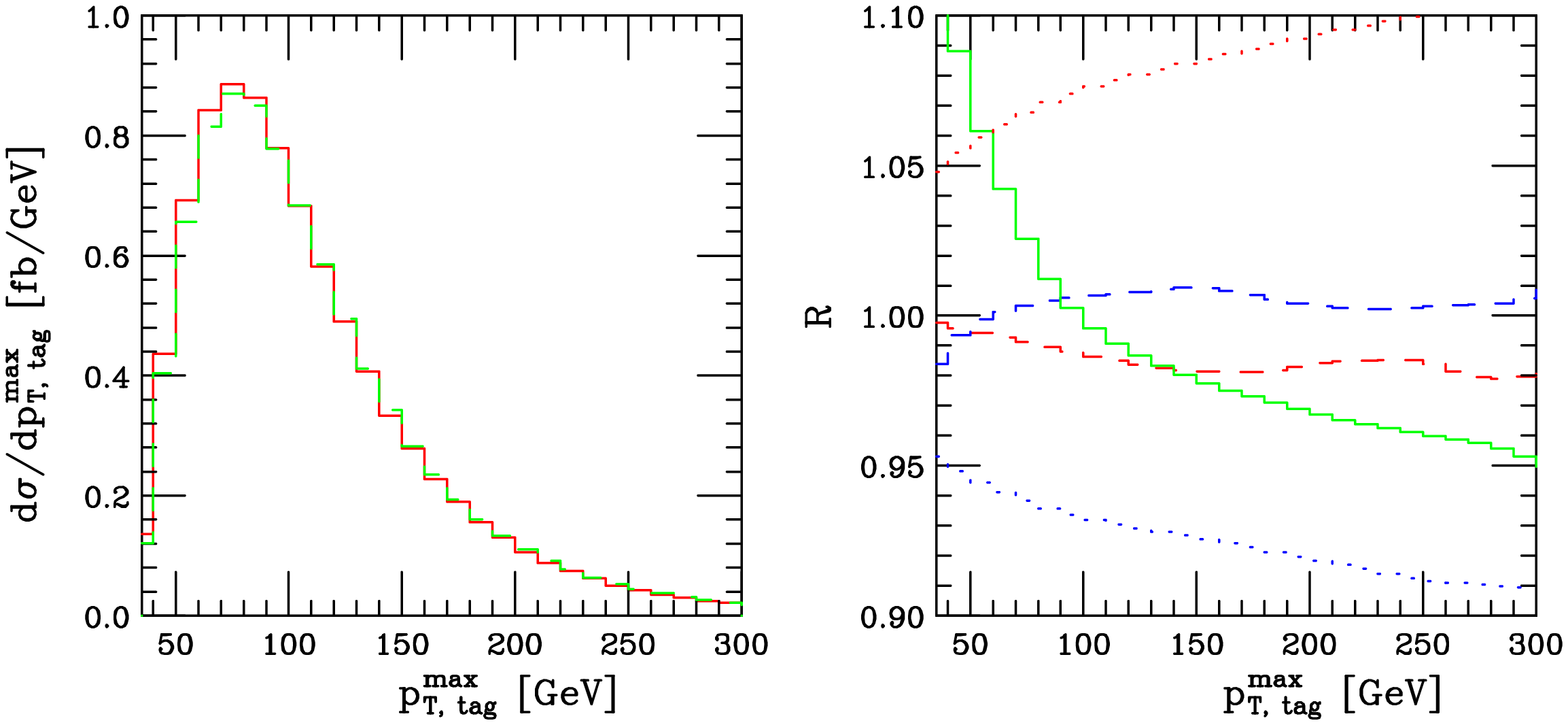,angle=90,width=4.5in}
\caption{Maximum transverse momentum of the two tagging jets.
In the left panel, $d\sigma/d p_{T,{\rm tag} }^{\rm max}$ is shown at LO (dashed green)
and NLO (solid red) for $\mu_0 = q_{V}$.  The right-hand 
panel depicts the $K$ factor (solid green) and scale variations of LO 
(dotted) and NLO (dashed) results for $\mu_R =\mu_F= \xi \mu_{0}$ 
with $\xi = 1/2$ and $2$.}%
\label{fig:ptmax}} 

\FIGURE{
\epsfig{file=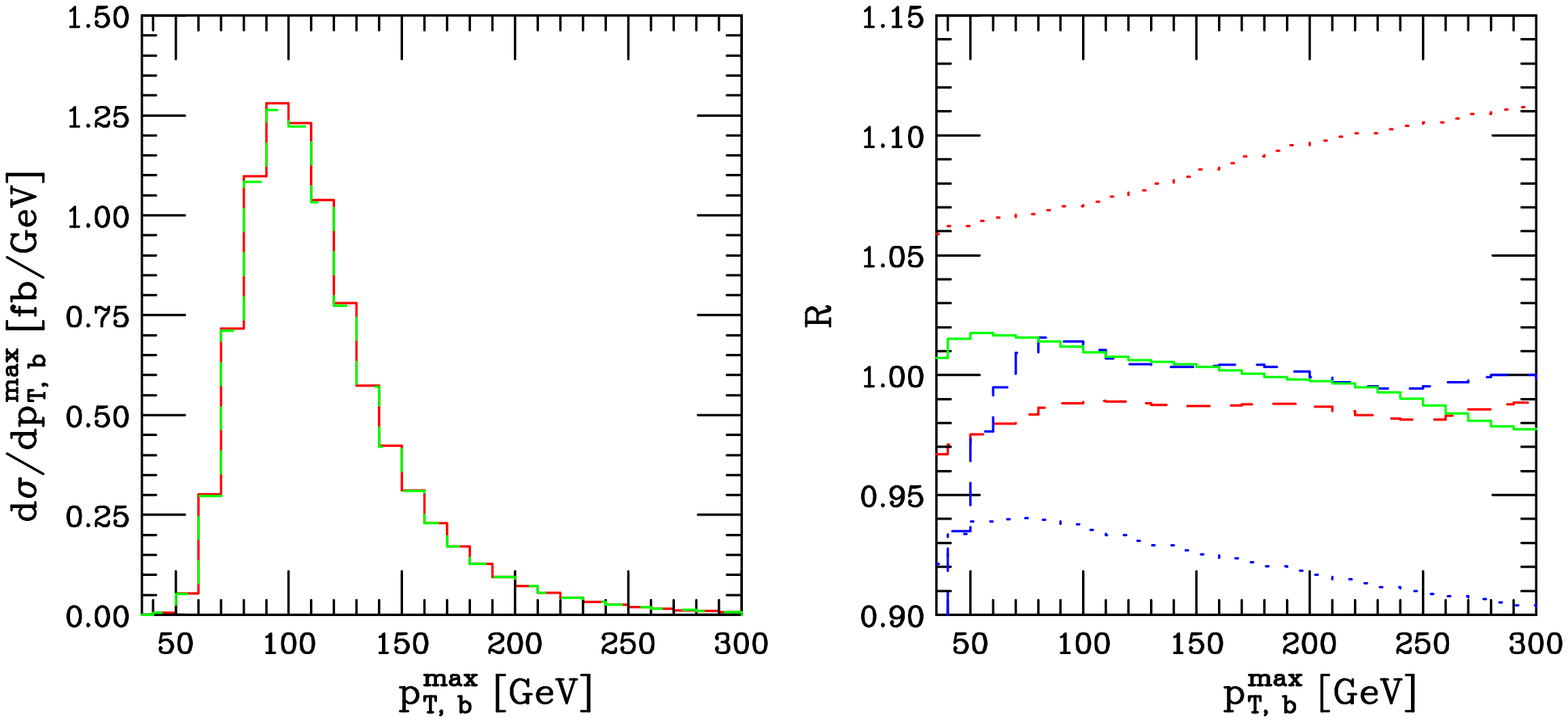,angle=90,width=4.5in}
\caption{Maximum transverse momentum of the four $b$ jets.
In the left panel, $d\sigma/d p_{T,{\rm b} }^{\rm max}$ is shown at LO (dashed green)
and NLO (solid red) for $\mu_0 = q_{V}$.  The right-hand 
panel depicts the $K$ factor (solid green) and scale variations of LO 
(dotted) and NLO (dashed) results for $\mu_R =\mu_F= \xi \mu_{0}$ 
with $\xi = 1/2$ and $2$.}%
\label{fig:ptbmax}}
 
\FIGURE{
\epsfig{file=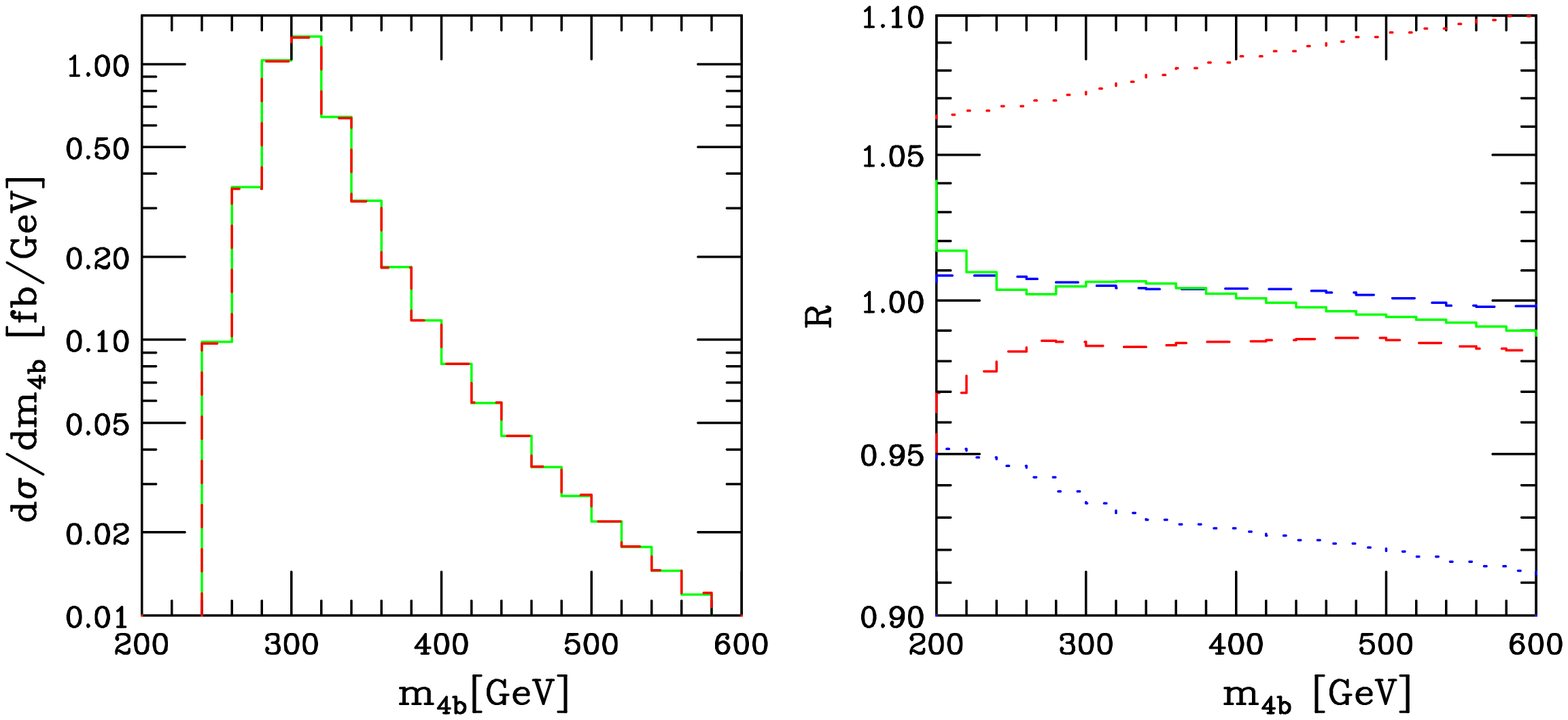,angle=90,width=4.5in}
\caption{The four $b$--jet invariant mass, $m_{4b}^{2} = (p_{b,1}+p_{b,2}+p_{b,3}+p_{b,4})^2$.
In the left panel, $d\sigma/d m_{4b}$ is shown at LO (dashed green)
and NLO (solid red) for $\mu_0 = q_{V}$.  The right-hand 
panel depicts the $K$ factor (solid green) and scale variations of LO 
(dotted) and NLO (dashed) results for $\mu_R =\mu_F= \xi \mu_{0}$ 
with $\xi = 1/2$ and $2$.}
\label{fig:m4b}
} 

\FIGURE{
\epsfig{file=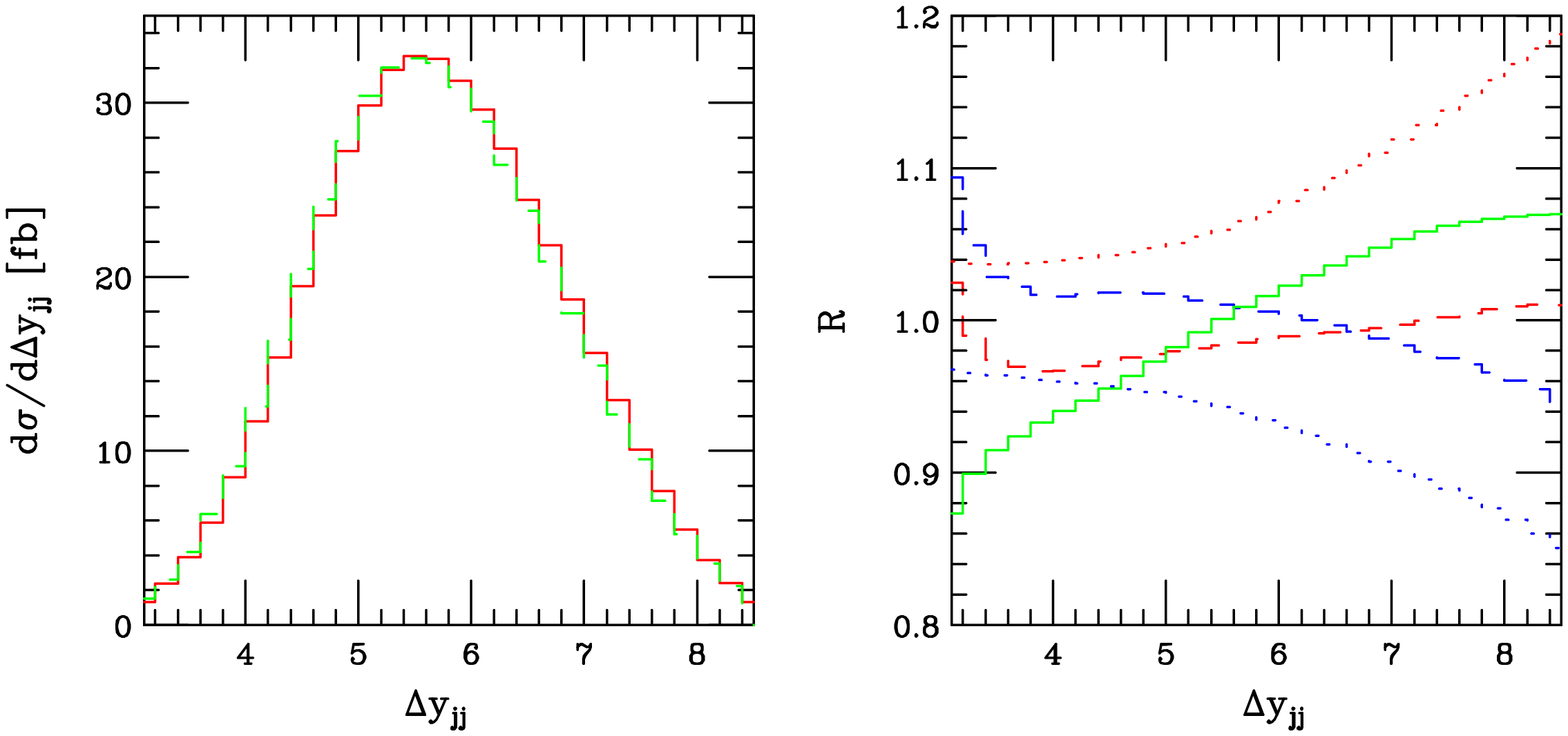,angle=90,width=4.5in}
\caption{Rapidity separation within the cuts of 
Eqs.~(\ref{cuts1})-(\ref{eq:dijet}). 
In the left panel, $d\sigma/d y_{jj}$ is shown at LO (dashed green)
and NLO (solid red) for $\mu_0 = q_{V}$.  The right-hand 
panel depicts the $K$ factor (solid green) and scale variations of LO 
(dotted) and NLO (dashed) results for $\mu_R =\mu_F= \xi \mu_{0}$ 
with $\xi = 1/2$ and $2$.}
\label{fig:yjj}
} 
\FIGURE{
\epsfig{file=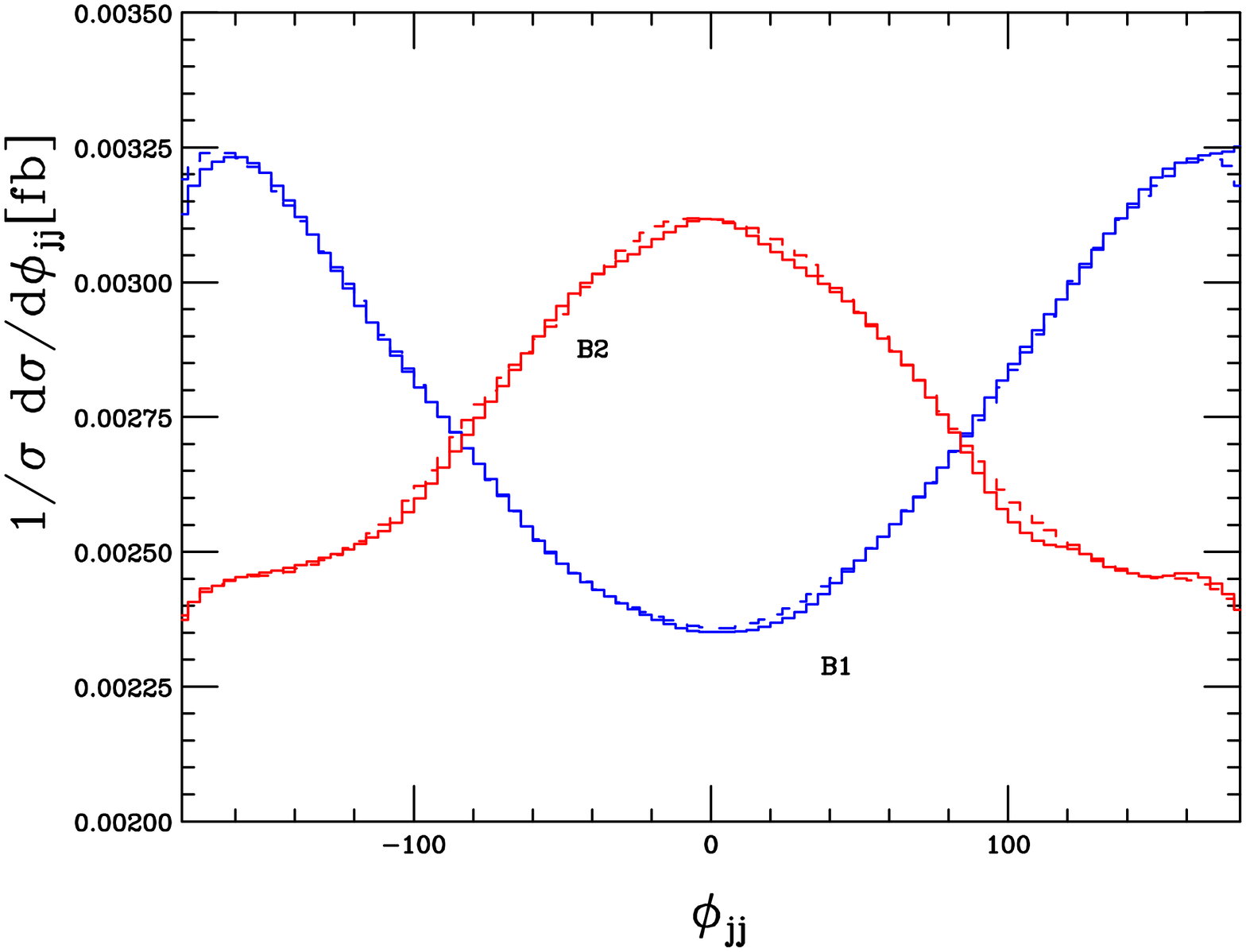,angle=90,width=3.8in}
\caption{The normalized distibution in azimuthal angle correlation of the two tagging jets,
$\frac{1}{\sigma}\frac{d \sigma}{d \Delta \phi_{jj}}$, for benchmark scenarios $B1$ (red) and $B2$ (blue) at NLO (solid) and LO (dashed).}%
\label{fig:azim}
} 

In Figure \ref{fig:azim}, we show the normalized azimuthal angle correlation of the tagging jets, $\Delta \phi_{jj}$, for benchmark points $B1$ and $B2$. Here $\Delta \phi_{jj}=\phi_{j_{+}}-\phi_{j_{-}}$ with $j_{+}$($j_{-}$)  
being the ``toward''(``away'') jet as defined in Ref.~\cite{Hankele:2006ma}.
For benchmark $B1$, the normalized 
distribution in $\Delta \phi_{jj}$ has the characteristic shape for single $H^{0}$ production followed by the decay, $H^{0} \to h^{0} h^{0}$ where the tensor structure of the $HVV$ vertex is $g^{\mu \nu}$ \cite{Hankele:2006ma,Figy:2004pt}.  However, for benchmark $B2$ in which the $H^{0}$ is decoupled from the gauge bosons the shape of the $\Delta \phi_{jj}$ distribution develops a peak at $\Delta \phi_{jj}=0$ degrees as opposed to a dip.\footnote{Such features have been pointed out in Ref.~\cite{Konar:2006qx} for
slepton pair production via VBF.} This is due to the interference of $s$--channel and $t$--channel $VV \to h^{0} h^{0}$ graphs. 
\section{Conclusions}
\label{sec:concl}

We have computed the next--to--leading order QCD corrections for light Higgs pair production via vector boson fusion at the LHC within the type II CP conserving two--higgs doublet model. We have included the subsequent decay of the light Higgs to the $b \bar{b}$ final state. Our NLO calculation takes the form of a fully flexible partonic Monte Carlo program allowing arbitrary phase space cuts. We have shown that scale variations for total cross sections and distributions are reduced at NLO. QCD $K$--factors are modest. These results are consistent with those of Standard Model Higgs production via vector boson fusion \cite{Figy:2003nv}. 

We also note the sensitivity of the azimuthal angle correlation of the two tagging jets, $ d \sigma / d \Delta\phi_{jj}$, to the tensorial structure of $VV \to h^{0} h^{0}$ scattering amplitudes.  For THDM scenarios in which the heavy Higgs coupling to electroweak gauge bosons is highly suppressed, the $\Delta \phi_{jj}$ distribution is peaked at $\Delta \phi_{jj}=0$ degrees while for scenarios in heavy Higgs couples to electroweak gauge bosons there is a dip at $\Delta \phi_{jj}=0$ degrees.

\acknowledgments{Work supported in part by the European Community's Marie-Curie Research
Training Network under contract MRTN-CT-2006-035505
`Tools and Precision Calculations for Physics Discoveries at Colliders'.  TF would like to thank 
Oliver Brein and Dieter Zeppenfeld for discussions concerning this project. All numerical computations presented in
this paper were performed via PhenoGrid using GridPP infrastructure.}

\end{document}